\input harvmac
%\draftmode
%\def\IR{\relax{\rm I\kern-.18em R}}
%\input epsf
%

\let\includefigures=\iftrue
\let\useblackboard=\iftrue
\newfam\black

%Figure Stuff
\includefigures
\message{If you do not have epsf.tex (to include figures),}
\message{change the option at the top of the tex file.}
\input epsf
\def\figin{\epsfcheck\figin}\def\figins{\epsfcheck\figins}
\def\epsfcheck{\ifx\epsfbox\UnDeFiNeD
\message{(NO epsf.tex, FIGURES WILL BE IGNORED)}
\gdef\figin##1{\vskip2in}\gdef\figins##1{\hskip.5in}% blank space instead
\else\message{(FIGURES WILL BE INCLUDED)}%
\gdef\figin##1{##1}\gdef\figins##1{##1}\fi}
\def\DefWarn#1{}
\def\figinsert{\goodbreak\midinsert}
\def\ifig#1#2#3{\DefWarn#1\xdef#1{fig.~\the\figno}
\writedef{#1\leftbracket fig.\noexpand~\the\figno}%
\figinsert\figin{\centerline{#3}}\medskip\centerline{\vbox{
\baselineskip12pt\advance\hsize by -1truein
\noindent\footnotefont{\bf Fig.~\the\figno:} #2}}
%\bigskip
\endinsert\global\advance\figno by1}
%%%
\else
\def\ifig#1#2#3{\xdef#1{fig.~\the\figno}
\writedef{#1\leftbracket fig.\noexpand~\the\figno}%
%\figinsert\figin{\centerline{#3}}\medskip
%\centerline{\vbox{\baselineskip12pt
%\advance\hsize by -1truein\noindent
%\footnotefont{\bf Fig.~\the\figno:} #2}}
%\bigskip\endinsert
\global\advance\figno by1} \fi

\def\journal#1&#2(#3){\unskip, \sl #1\ \bf #2 \rm(19#3) }
\def\andjournal#1&#2(#3){\sl #1~\bf #2 \rm (19#3) }

\def\ie{{\it i.e.}}
\def\eg{{\it e.g.}}

\noblackbox
%

% Something to deal with sub-sub-sections

\def\unlockat{\catcode`\@=11}
\def\lockat{\catcode`\@=12}

\unlockat
% Something to deal with sub-sub-sections

\def\newsec#1{\global\advance\secno by1\message{(\the\secno. #1)}
\global\subsecno=0\global\subsubsecno=0\eqnres@t\noindent
{\bf\the\secno. #1}
\writetoca{{\secsym} {#1}}\par\nobreak\medskip\nobreak}
\global\newcount\subsecno \global\subsecno=0
\def\subsec#1{\global\advance\subsecno
by1\message{(\secsym\the\subsecno. #1)}
\ifnum\lastpenalty>9000\else\bigbreak\fi\global\subsubsecno=0
\noindent{\it\secsym\the\subsecno. #1}
\writetoca{\string\quad {\secsym\the\subsecno.} {#1}}
\par\nobreak\medskip\nobreak}
\global\newcount\subsubsecno \global\subsubsecno=0
\def\subsubsec#1{\global\advance\subsubsecno by1
\message{(\secsym\the\subsecno.\the\subsubsecno. #1)}
\ifnum\lastpenalty>9000\else\bigbreak\fi
\noindent\quad{\secsym\the\subsecno.\the\subsubsecno.}{#1}
\writetoca{\string\qquad{\secsym\the\subsecno.\the\subsubsecno.}{#1}}
\par\nobreak\medskip\nobreak}

\def\subsubseclab#1{\DefWarn#1\xdef
#1{\noexpand\hyperref{}{subsubsection}%
{\secsym\the\subsecno.\the\subsubsecno}%
{\secsym\the\subsecno.\the\subsubsecno}}%
\writedef{#1\leftbracket#1}\wrlabeL{#1=#1}}% Macros for boxes
\lockat

\def\ie{{\it i.e.}}
\def\eg{{\it e.g.}}

%% MORE MACROS
\def\CM {{\cal M}}
\def\CN {{\cal N}}

\def\CO {{\cal O}}

\def\CS {{\cal S }}

\font\manual=manfnt \def\dbend{\lower3.5pt\hbox{\manual\char127}}

\def\IZ{\relax\ifmmode\mathchoice
{\hbox{\cmss Z\kern-.4em Z}}{\hbox{\cmss Z\kern-.4em Z}}
{\lower.9pt\hbox{\cmsss Z\kern-.4em Z}}
{\lower1.2pt\hbox{\cmsss Z\kern-.4em Z}}\else{\cmss Z\kern-.4em
Z}\fi}
\def\half{{1\over 2}}

\def\CM {{\cal M}}
\def\CN {{\cal N}}

\def\CO {{\cal O}}

\def\CS {{\cal S }}

% more macros, alphabetically

\def\IZ{\relax\ifmmode\mathchoice
{\hbox{\cmss Z\kern-.4em Z}}{\hbox{\cmss Z\kern-.4em Z}}
{\lower.9pt\hbox{\cmsss Z\kern-.4em Z}}
{\lower1.2pt\hbox{\cmsss Z\kern-.4em Z}}\else{\cmss Z\kern-.4em
Z}\fi}
\def\IB{\relax{\rm I\kern-.18em B}}
\def\IC{{\relax\hbox{$\inbar\kern-.3em{\rm C}$}}}
\def\ID{\relax{\rm I\kern-.18em D}}
\def\IE{\relax{\rm I\kern-.18em E}}
\def\IF{\relax{\rm I\kern-.18em F}}
\def\IG{\relax\hbox{$\inbar\kern-.3em{\rm G}$}}
\def\IGa{\relax\hbox{${\rm I}\kern-.18em\Gamma$}}
\def\IH{\relax{\rm I\kern-.18em H}}
\def\II{\relax{\rm I\kern-.18em I}}
\def\IK{\relax{\rm I\kern-.18em K}}
\def\IP{\relax{\rm I\kern-.18em P}}
\def\IQ{\relax\hbox{$\inbar\kern-.3em{\rm Q}$}}

\def\inbar{\,\vrule height1.5ex width.4pt depth0pt}

\font\cmss=cmss10 \font\cmsss=cmss10 at 7pt
\def\IR{\relax{\rm I\kern-.18em R}}

% Macros for boxes
%
%\def\boxit#1{\vbox{\hrule\hbox{\vrule\kern8pt
%\vbox{\hbox{\kern8pt}\hbox{\vbox{#1}}\hbox{\k
%\hbox{$\displaystyle #1$}\kern8pt}\kern8pt\vrule}\hrule}}}
%
%%% MACROS FOR BOX BOUNDARY CONDS
%%% FROM KAWAI ET AL

\def\makeblankbox#1#2{\hbox{\lower\dp0\vbox{\hidehrule{#1}{#2}%
   \kern -#1% overlap rules
   \hbox to \wd0{\hidevrule{#1}{#2}%
      \raise\ht0\vbox to #1{}% vrule height
      \lower\dp0\vtop to #1{}% vrule depth
      \hfil\hidevrule{#2}{#1}}%
   \kern-#1\hidehrule{#2}{#1}}}%
}%
\def\hidehrule#1#2{\kern-#1\hrule height#1 depth#2 \kern-#2}%
\def\hidevrule#1#2{\kern-#1{\dimen0=#1\advance\dimen0 by #2\vrule
    width\dimen0}\kern-#2}%
\def\openbox{\ht0=1.2mm \dp0=1.2mm \wd0=2.4mm  \raise 2.75pt
\makeblankbox {.25pt} {.25pt}  }

\def\bun#1/#2{\leavevmode
   \kern.1em \raise .5ex \hbox{\the\scriptfont0 #1}%
   \kern-.1em $/$%
   \kern-.15em \lower .25ex \hbox{\the\scriptfont0 #2}%
}

\def\opensquare{\ht0=3.4mm \dp0=3.4mm \wd0=6.8mm  \raise 2.7pt
\makeblankbox {.25pt} {.25pt}  }

%%%%%%%%%%%%%%%%%%%%%%%

\def\sector#1#2{\ {\scriptstyle #1}\hskip 1mm
\mathop{\opensquare}\limits_{\lower 1mm\hbox{$\scriptstyle#2$}}\hskip 1mm}

\def\tsector#1#2{\ {\scriptstyle #1}\hskip 1mm
\mathop{\opensquare}\limits_{\lower 1mm\hbox{$\scriptstyle#2$}}^\sim\hskip 1mm}
%%%
%%%

%% ANOTHER SET OF MACROS

\def\inbar{\,\vrule height1.5ex width.4pt depth0pt}

\font\cmss=cmss10 \font\cmsss=cmss10 at 7pt
\def\IR{\relax{\rm I\kern-.18em R}}

%% new macros

\def\frac#1#2{{#1\over#2}}

\def\half{\frac12}

\def\inbar{\,\vrule height1.5ex width.4pt depth0pt}
\def\IC{\relax\hbox{$\inbar\kern-.3em{\rm C}$}}
\def\IR{\relax{\rm I\kern-.18em R}}
\def\IP{\relax{\rm I\kern-.18em P}}

%
%%%%%%%%%%%%%%%%%%%%%%%%%%%%%%%%%%%%
%
\catcode`\@=11
\def\slash#1{\mathord{\mathpalette\c@ncel{#1}}}
\overfullrule=0pt

\def\II{{\cal I}}

\def\underrel#1\over#2{\mathrel{\mathop{\kern\z@#1}\limits_{#2}}}

\catcode`\@=12

%%%%%%%%%%%%%%%%%%%%%%%%%%%%%%%%%%%%%%%%%%%%%%%%%%%%%%%%%%%%%%

%

\def\exp{{\rm exp}}

%%%%%%%%%%%%%%%%%%%%%%%%%%%%%%%%%%%%%%%%%%%%%%%%%%%%%%%%%%%%%%
% new defs:

%%%%%%%%%%%%%%%%%%%%%%%%%%%%%%%%%%%%%%%%%%%%%%%%%%%%%%%%%%%%%%

\def\frac#1#2{{#1\over#2}}

\def\half{\frac12}

\def\inbar{\,\vrule height1.5ex width.4pt depth0pt}
\def\IC{\relax\hbox{$\inbar\kern-.3em{\rm C}$}}
\def\IR{\relax{\rm I\kern-.18em R}}
\def\IP{\relax{\rm I\kern-.18em P}}

%
%%%%%%%%%%%%%%%%%%%%%%%%%%%%%%%%%%%%
%

%
\catcode`\@=11
\def\slash#1{\mathord{\mathpalette\c@ncel{#1}}}
\overfullrule=0pt

\def\II{{\cal I}}

\def\underrel#1\over#2{\mathrel{\mathop{\kern\z@#1}\limits_{#2}}}

\catcode`\@=12

%%%%%%%%%%%%%%%%%%%%%%%%%%%%%%%%%%%%%%%%%%%%%%%%%%%%%%%%%%%%%%

%

\def\exp{{\rm exp}}

%%%%%%%%%%%%%%%%%%%%%%%%%%%%%%%%%%%%%%%%%%%%%%%%%%%%%%%%%%%%%%
% new defs:

%%%%%%%%%%%%%%%%%%%%%%%%%%%%%%%%%%%%%%%%%%%%%%%%%%%%%%%%%%%%%%%%%%%%%%%%%%%%%%%%%%

%\CalabreseQY
\lref\CalabreseQY{
  P.~Calabrese and J.~Cardy,
  ``Entanglement entropy and conformal field theory,''
J.\ Phys.\ A {\bf 42}, 504005 (2009).
[arXiv:0905.4013 [cond-mat.stat-mech]].
%%CITATION = arXiv:0905.4013%%
}

%\RyuEF
\lref\RyuEF{
  S.~Ryu and T.~Takayanagi,
  ``Aspects of Holographic Entanglement Entropy,''
JHEP {\bf 0608}, 045 (2006).
[hep-th/0605073].
%%CITATION = hep-th/0605073%%
}

%\DixonQV
\lref\DixonQV{
  L.~J.~Dixon, D.~Friedan, E.~J.~Martinec and S.~H.~Shenker,
  ``The Conformal Field Theory of Orbifolds,''
Nucl.\ Phys.\ B {\bf 282}, 13 (1987).
%%CITATION = EFI-86-42-CHICAGO%%
}

%\ArgurioTB
\lref\ArgurioTB{
  R.~Argurio, A.~Giveon and A.~Shomer,
  ``Superstrings on AdS(3) and symmetric products,''
JHEP {\bf 0012}, 003 (2000).
[hep-th/0009242].
%%CITATION = hep-th/0009242%%
}

%\KnizhnikKF
\lref\KnizhnikKF{
  V.~G.~Knizhnik,
  ``Analytic Fields On Riemannian Surfaces,''
Phys.\ Lett.\ B {\bf 180}, 247 (1986).
%%CITATION = Print-86-0908 (LANDAU)%%
}

%\KnizhnikXP
\lref\KnizhnikXP{
  V.~G.~Knizhnik,
  ``Analytic Fields on Riemann Surfaces. 2,''
Commun.\ Math.\ Phys.\  {\bf 112}, 567 (1987).
}

%\KlemmDF
\lref\KlemmDF{
  A.~Klemm and M.~G.~Schmidt,
  ``Orbifolds by Cyclic Permutations of Tensor Product Conformal Field Theories,''
Phys.\ Lett.\ B {\bf 245}, 53 (1990).
%%CITATION = HD-THEP-90-13%%
}

%\FuchsVU
\lref\FuchsVU{
  J.~Fuchs, A.~Klemm and M.~G.~Schmidt,
  ``Orbifolds by cyclic permutations in Gepner type superstrings and in the corresponding Calabi-Yau manifolds,''
Annals Phys.\  {\bf 214}, 221 (1992).
%%CITATION = HD-THEP-90-34%%
}

%\GiveonNS
\lref\GiveonNS{
  A.~Giveon, D.~Kutasov and N.~Seiberg,
  ``Comments on string theory on AdS(3),''
Adv.\ Theor.\ Math.\ Phys.\  {\bf 2}, 733 (1998).
[hep-th/9806194].
%%CITATION = hep-th/9806194%%
}

%\KutasovZH
\lref\KutasovZH{
  D.~Kutasov, F.~Larsen and R.~G.~Leigh,
  ``String theory in magnetic monopole backgrounds,''
Nucl.\ Phys.\ B {\bf 550}, 183 (1999).
[hep-th/9812027].
%%CITATION = hep-th/9812027%%
}

%\GiveonJG
\lref\GiveonJG{
  A.~Giveon and M.~Rocek,
  ``Supersymmetric string vacua on AdS(3) x N,''
JHEP {\bf 9904}, 019 (1999).
[hep-th/9904024].
%%CITATION = hep-th/9904024%%
}

%\BerensteinGJ
\lref\BerensteinGJ{
  D.~Berenstein and R.~G.~Leigh,
  ``Space-time supersymmetry in AdS(3) backgrounds,''
Phys.\ Lett.\ B {\bf 458}, 297 (1999).
[hep-th/9904040].
%%CITATION = hep-th/9904040%%
}

%\MaldacenaHW
\lref\MaldacenaHW{
  J.~M.~Maldacena and H.~Ooguri,
  ``Strings in AdS(3) and SL(2,R) WZW model 1.: The Spectrum,''
J.\ Math.\ Phys.\  {\bf 42}, 2929 (2001).
[hep-th/0001053].
%%CITATION = hep-th/0001053%%
}

%\KutasovXU
\lref\KutasovXU{
  D.~Kutasov and N.~Seiberg,
  ``More comments on string theory on AdS(3),''
JHEP {\bf 9904}, 008 (1999).
[hep-th/9903219].
%%CITATION = hep-th/9903219%%
}

%\NishiokaUN
\lref\NishiokaUN{
  T.~Nishioka, S.~Ryu and T.~Takayanagi,
  ``Holographic Entanglement Entropy: An Overview,''
J.\ Phys.\ A {\bf 42}, 504008 (2009).
[arXiv:0905.0932 [hep-th]].
%%CITATION = arXiv:0905.0932%%
}

%\PorratiEHA
\lref\PorratiEHA{
  J.~Kim and M.~Porrati,
  ``On the central charge of spacetime current algebras and correlators in string theory on AdS$_{3}$,''
JHEP {\bf 1505}, 076 (2015).
[arXiv:1503.07186 [hep-th]].
%%CITATION = arXiv:1503.07186%%
}

%\NishiokaHAA
\lref\NishiokaHAA{
  T.~Nishioka and I.~Yaakov,
  ``Supersymmetric Renyi Entropy,''
JHEP {\bf 1310}, 155 (2013).
[arXiv:1306.2958 [hep-th]].
%%CITATION = PUPT-2448%%
}

%\HuangPDA
\lref\HuangPDA{
  X.~Huang and Y.~Zhou,
  ``$ \CN=4 $ Super-Yang-Mills on conic space as hologram of STU topological black hole,''
JHEP {\bf 1502}, 068 (2015).
[arXiv:1408.3393 [hep-th]].
%%CITATION = arXiv:1408.3393%%
}

%\CrossleyOEA
\lref\CrossleyOEA{
  M.~Crossley, E.~Dyer and J.~Sonner,
  ``Super-Renyi entropy \& Wilson loops for $ \CN=4 $ SYM and their gravity duals,''
JHEP {\bf 1412}, 001 (2014).
[arXiv:1409.0542 [hep-th]].
%%CITATION = MIT-CTP-4579%%
}

%\HamaIEA
\lref\HamaIEA{
  N.~Hama, T.~Nishioka and T.~Ugajin,
  ``Supersymmetric Renyi entropy in five dimensions,''
JHEP {\bf 1412}, 048 (2014).
[arXiv:1410.2206 [hep-th]].
%%CITATION = YITP-14-78%%
}

%\NishiokaMWA
\lref\NishiokaMWA{
  T.~Nishioka,
  ``The Gravity Dual of Supersymmetric Renyi Entropy,''
JHEP {\bf 1407}, 061 (2014).
[arXiv:1401.6764 [hep-th]].
%%CITATION = arXiv:1401.6764%%
}

%\AldayFSA
\lref\AldayFSA{
  L.~F.~Alday, P.~Richmond and J.~Sparks,
  ``The holographic supersymmetric Renyi entropy in five dimensions,''
JHEP {\bf 1502}, 102 (2015).
[arXiv:1410.0899 [hep-th]].
%%CITATION = arXiv:1410.0899%%
}

%\HuangGCA
\lref\HuangGCA{
  X.~Huang, S.~J.~Rey and Y.~Zhou,
  ``Three-dimensional SCFT on conic space as hologram of charged topological black hole,''
JHEP {\bf 1403}, 127 (2014).
[arXiv:1401.5421 [hep-th]].
%%CITATION = arXiv:1401.5421%%
}

%%%%%%%%%%%%%%%%%%%%%%%%%%%%%%%%%%%%%%%%%%%%%%%%%%%%%%%%%%%%%%%%%%%%%%%%%%%%%
%\rightline{....}
\Title{
%\rightline{hep-th/yymmnnn}
} {\vbox{
\bigskip\centerline{Supersymmetric Renyi Entropy in $CFT_2$ and $AdS_3$}}}
\medskip
\centerline{\it Amit Giveon${}^{1}$ and David Kutasov${}^{2}$}
\bigskip
\smallskip
\centerline{${}^{1}$Racah Institute of Physics, The Hebrew
University} \centerline{Jerusalem 91904, Israel}
\smallskip
\centerline{${}^2$EFI and Department of Physics, University of
Chicago} \centerline{5640 S. Ellis Av., Chicago, IL 60637, USA }

\bigskip\bigskip\bigskip
\noindent

We show that in any two dimensional conformal field theory with $(2,2)$ supersymmetry one can define a supersymmetric analog of the usual Renyi entropy of a spatial region $A$. It differs from the Renyi entropy by a universal function (which we compute) of the central charge,  Renyi parameter $n$ and the geometric parameters of $A$. In the limit $n\to 1$ it coincides with the entanglement entropy. Thus, it contains the same information as the Renyi entropy but its computation only involves correlation functions of chiral and anti-chiral operators. We also show that this quantity appears naturally in string theory on $AdS_3$.

\vglue .3cm
%\vskip 2cm
\bigskip

\Date{10/15}

\bigskip

\newsec{Introduction}

In recent years there has been a lot of work on spatial entanglement in quantum field theory; see \eg\ \refs{\NishiokaUN,\CalabreseQY} for reviews. In a prototypical construction one divides space into two complementary regions, $A$ and $B$, and traces the density matrix of the vacuum,\foot{One can replace the vacuum state $|0\rangle$ by any other pure or mixed state in this construction, but we will not do that here.} $\rho_0=|0\rangle\langle0|$, over the degrees of freedom in region $B$, thus forming the reduced density matrix
\eqn\aaa{\rho_A={\rm Tr}_B\rho_0~.}
This density matrix contains information about the entanglement between the degrees of freedom in regions $A$ and $B$ in the state $|0\rangle$. A convenient way to access this information is via the Renyi  entropy
\eqn\bbb{S_A^{(n)}={1\over 1-n}\ln\Tr\rho_A^n~,}
where $n$ is initially a positive integer. One can often continue \bbb\ to arbitrary real $n$, and in particular take the limit $n\to 1$, which gives the entanglement entropy
\eqn\ccc{S_A=\lim_{n\to 1} S_A^{(n)}=-{\rm Tr}\rho_A\ln\rho_A~.}
This is known as the replica trick.

In general, calculating the entropies \bbb, \ccc\ is hard, even for free field theories. However, in $1+1$ dimensional conformal field theory (CFT)  the problem simplifies somewhat. One can take the region $A$ to be the union of $N$ (ordered) disjoint intervals
\eqn\ddd{A=(u_1,v_1)\cup(u_2,v_2)\cdots\cup (u_N,v_N)~, }
and $B$ its complement in $\IR$. The problem of calculating the Renyi entropy \bbb\ for integer $n$ can then be mapped to one involving correlation functions of certain twist fields $\CT_n$. These twist fields can be thought of as living in the symmetric product CFT
\eqn\symprod{\CM^p/S_p~,}
where $\CM$ is the original CFT with central charge $c$, and $p\ge n$ is an arbitrary integer. The operator $\CT_n$ is the lowest dimension operator in a $\IZ_n$ twisted sector.  The generator of $\IZ_n$ takes the $i$'th copy of $\CM$ to the $(i+1)$'th, with the $n$'th copy mapped back to the first. The dimension of $\CT_n$ is given by
\refs{\KnizhnikKF\KnizhnikXP\KlemmDF-\FuchsVU}
\eqn\eee{h_n=\bar h_n={c\over24}\left(n-{1\over n}\right)~.}
One can show \CalabreseQY\ that the Renyi entropy \bbb\ is given\foot{Up to an overall constant that will not play a role in our discussion.} by
\eqn\fff{\Tr\rho_A^n=\langle \CT_n(u_1)\CT_{n}^*(v_1)\CT_n(u_2)\CT_{n}^*(v_2)\cdots \CT_n(u_N)\CT_{n}^*(v_N)\rangle~,}
where all fields are inserted at a particular time $t$, on which nothing depends.

{}For one interval ($N=1$, $u_1=u$, $v_1=v$, $l=|v-u|$), one has
\eqn\ggg{\Tr\rho_A^n=\langle \CT_n(u)\CT_{n}^*(v)\rangle\sim l^{-4h_n}~,}
which leads to
\eqn\oneint{S_A^{(n)}={c\over 6}\left(1+{1\over n}\right)\ln{l\over a}~;\qquad S_A={c\over 3}\ln{l\over a}~,}
with $a$ an arbitrary constant. For $N>1$ intervals one has to evaluate $2N$ point functions of twist fields, which is in general more involved, but can in principle be done using orbifold CFT techniques.

The construction above is general, but it is natural to ask whether it simplifies in CFT's with additional symmetries. In this note, we will discuss this question for the case where the additional symmetry is supersymmetry. $\CN=1$ supersymmetry usually does not buy one much in $2d$ CFT, but theories with $\CN=2$ superconformal symmetry do enjoy many special properties. In particular, the $\CN=2$ superconformal algebra contains a conserved $U(1)_R$ current,  and there is a special class of operators, known as chiral operators, that belong to short multiplets, and have the property that their scaling dimensions $h$ and R-charges $R$ satisfy the relation\foot{There are also anti-chiral operators that satisfy the opposite relation $h=-\half R$.}
\eqn\hhh{h=\half R~.}
Chiral operators form a ring -- the chiral ring -- which controls many properties of their correlation functions.

The operator $\CT_n$ that figures in the discussion of the Renyi entropy \fff\ is not chiral. In fact, as we review below, the orbifold that enters the construction of \CalabreseQY\ is in general not supersymmetric, even if the theory $\CM$ is. This is analogous to the fact that when one computes the thermal free energy for a supersymmetric field theory by putting Euclidean time on a circle, supersymmetry is broken by the boundary conditions for fermions on the circle.

In this note we will show that the information contained in the correlation function \fff\ can be obtained by studying a supersymmetric analog of \symprod, in which the role of the operator $\CT_n$ is played by a certain closely related operator $\CS_n$, which is chiral. The analog of \fff\ is a correlation function of $N$ chiral operators $\CS_n$ inserted at the points $u_i$, and $N$ anti-chiral operators $\CS_n^*$ inserted at $v_i$. We will refer to the resulting $2N$ point function as {\it supersymmetric Renyi entropy} (SRE), in analogy to \fff. We will show that it differs from its non-supersymmetric analog \fff\ by a universal function of the parameters $(c, n, u_i,v_i)$ and thus contains the same information. At the same time it might be easier to compute by using the special properties of chiral operators in $\CN=2$ SCFT. 

Another advantage of the supersymmetric Renyi entropy is that in $(2,2)$ SCFT's that have an $AdS_3$ dual, it can be calculated in the bulk. In string theory on $AdS_3$ with Neveu-Schwarz $B$ field, the operators $\CS_n$ are described by physical vertex operators, and thus the  SRE can be calculated in the full string theory rather than in the gravity limit, as in \NishiokaUN.

In the rest of this note we describe the construction of the chiral operators $\CS_n$, and the relation of their correlation functions which give the supersymmetric Renyi entropy to the usual Renyi entropy \fff. In section 2 we study the special case where the CFT $\CM$ consists of one free chiral superfield. In section 3 we extend the discussion to a general $(2,2)$ SCFT. In section 4 we discuss the AdS/CFT dual of SRE. Section 5 contains some comments.

\newsec{Warm-up exercise: a single free chiral superfield}

In this section we consider the case where the  SCFT $\CM$, whose Renyi entropy we would like to compute, consists of one free chiral superfield. The bottom component of this superfield is a complex scalar $\phi$; its superpartner under the left (right) moving $\CN=2$ supersymmetry is a complex left (right) moving fermion $\psi$ $(\bar\psi)$. The left-moving $\CN=2$ superconformal generators take the form $G^+\sim \psi\partial\phi^*$, $G^-\sim \psi^*\partial\phi$. The central charge of this theory is $c=3$. The structure of the right-moving sector is similar.

We start by reviewing the results of \CalabreseQY\ for this case. To construct the twist field $\CT_n$, we start with $n$ copies of the SCFT $\CM$, \ie\ $n$ chiral superfields\foot{Here and below we will mostly focus on the left-movers; the right-moving sector will be added later.} $(\phi_j,\psi_j)$, $j=1,\cdots, n$ and consider the cyclic orbifold $\CM^n/\IZ_n$.  The generator of $\IZ_n$ takes $(\phi_j,\psi_j)\to(\phi_{j+1},\psi_{j+1})$. However, the scalars and fermions obey different periodicity conditions in $j$,
\eqn\jjj{\phi_{j+n}=\phi_j~,\qquad \psi_{j+n}=(-)^{n-1}\psi_j~.}
One can think of the label $j$ in \jjj\ as parametrizing a discretized circle, and of \jjj\ as specifying the periodicity of the fields around the circle. The bosons are always periodic, while the fermions are (anti) periodic for (even) odd $n$. In analogy to the case of the thermal free energy, we expect supersymmetry to be broken by the boundary conditions for even $n$, and to remain unbroken for odd $n$.

As is standard for $\IZ_n$ orbifolds, we can diagonalize the $\IZ_n$ action by a discrete Fourier transform,
\eqn\disfour{\eqalign{\tilde\phi_k=&{1\over\sqrt n}\sum_{j=1}^n\phi_j\exp{2\pi i jk\over n}~,\cr
\tilde\psi_k=&{1\over\sqrt n}\sum_{j=1}^n\psi_j
\exp{2\pi ij\over n}\left[k-\half(n-1)\right]~,\cr
}}
where $k=0,1,2,\cdots,n-1$ labels the discrete momentum. The bosonic and fermonic fields \disfour\ are multiplied  by a phase under $\IZ_n$,
\eqn\multphase{\eqalign{\tilde\phi_k\to & \tilde\phi_k e^{-{2\pi ik\over n}}~,\cr
\tilde\psi_k\to& \tilde\psi_k e^{-{2\pi i\over n}\left(k-\half(n-1)\right)}~.}}
The transformation properties \multphase\ can be thought of as due to an insertion of a twist field, which has the property that as the fields $\tilde\phi_k$, $\tilde\psi_k$ encircle it, they are multiplied by the appropriate phase. For the scalar field $\tilde\phi_k$, the corresponding twist field $\sigma_k$ has dimension given by
(see \eg\ \DixonQV)
\eqn\dimsigma{h(\sigma_k)=\bar h(\sigma_k)=\half{k\over n}\left(1-{k\over n}\right)~.}
To construct the fermionic twist fields, it is convenient to bosonize the fermions $\tilde\psi_k$, and write them as
\eqn\bospsi{\tilde\psi_k=e^{iH_k}~,}
where $H_k$ are canonically normalized left-moving scalar fields. In terms of $H_k$, the fermionic twist field for $\tilde\psi_k$ takes the form
\eqn\fermtwist{s_k=e^{{i\over n}\left(k-\half(n-1)\right)H_k}~.}
Its dimension is given by
\eqn\dimsk{h(s_k)={1\over 2n^2}\left(k-\half(n-1)\right)^2~.}
The operator $\CT_n$ discussed in section 1 takes in this case the form\foot{After adding the right-movers  we have
$\CT_n=\prod\sigma_k s_k\bar s_k$, where $\bar s_k$ are the right-moving analogs of $s_k$. }
\eqn\formttnn{\CT_n=\prod_{k=0}^{n-1}\sigma_k s_k~.}
Adding up the dimensions \dimsigma,  \dimsk, we find
\eqn\hhttnn{h(\CT_n)=\bar h(\CT_n)={n^2-1\over 8n}~,}
in agreement with \eee\ for $c=3$.

As mentioned above, we expect this construction to break supersymmetry for even $n$. Each of the $n$ copies of the SCFT $\CM$ labeled by $j$
contains superconformal generators $G^+_j\sim \psi_j\partial\phi^*_j$, $G^-_j\sim \psi_j^*\partial\phi_j$, but the total superconformal generators,
\eqn\diaggpm{G^\pm=\sum_{j=1}^n G_j^\pm~,}
do not have zero modes in this case. A quick way to see that is to note that in \disfour\ the momentum on the discretized circle is integer for the bosonic field $\phi$, while for the fermionic field $\psi$ it takes value in $\IZ-\half(n-1)$. Thus, for even $n$ it is half integer, so SUSY is broken by boundary conditions.  Equivalently, one notes that applying $G^+$ to  $\tilde\phi_k$ (first line of \disfour) gives $G^+\tilde\phi_k=\tilde\psi_{k+\half(n-1)}$. For even $n$, the latter is not one of the modes on the second line of \disfour. Note that the $U(1)_R$ current,
\eqn\uone{J=\sum_{j=1}^n \psi^*_j\psi_j=i\sum_{k=0}^{n-1} \partial H_k~,}
does survive the orbifold, but since for even $n$ there are no supercharges for it to act on, it gives rise to a standard $U(1)$ global symmetry of the orbifold theory. The twist field $\CT_n$ is not charged under this symmetry. 

So far we discussed the cyclic orbifold $\CM^n/\IZ_n$. In terms of the symmetric product CFT \symprod, one can view it as a  sector of that CFT, corresponding to a particular conjugacy class, $(n)(1)^{p-n}$. The orbifold \symprod\ has cycles with both even and odd $n$; hence, it is not supersymmetric when defined via \jjj.

To summarize, the calculation of the Renyi entropy for a single chiral superfield involves $2N$ point functions (where $N$ is the number of intervals, as above) of the operators \formttnn, which contain bosonic and fermionic twist fields. It is worth pointing out that the difficult part of the calculation is the one associated with the bosonic twist fields. The fermionic ones are written in terms of free fields \fermtwist, and thus their correlation functions can be computed exactly using standard methods.

We would like to modify the construction of \CalabreseQY\ such that the $\CN=2$ superconformal symmetry generated by \diaggpm\ remains unbroken. The discussion above makes it clear how to proceed. We need to modify the conditions \jjj\ such that the fermions and bosons have the same periodicity. Thus, we want the fermions to be periodic, $\psi_{j+n}=\psi_j$,  and their Fourier expansion should be the same as that of the bosons in \disfour.

The second line of \multphase\ is now modified to $\tilde\psi_k\to \tilde\psi_k e^{-{2\pi ik\over n}}$. To implement this we replace the twist field \fermtwist\ with
\eqn\susyfermtwist{\hat s_k=e^{{ik\over n}H_k}~,}
whose dimension is
\eqn\dimshatk{h(\hat s_k)=\half\left(k\over n\right)^2~.}
As before, the index $k$ goes over the range $k=0,1,2,\cdots, n-1$. Note that for $n>k>n/2$ the operator \susyfermtwist\ is not the lowest dimension operator which implements the right twist for the fermion $\tilde\psi_k$. That operator is obtained by taking $k\to k-n$ in \susyfermtwist. However, $\hat s_k$ has the advantage that the modified twist field,
\eqn\formssnn{\CS_n=\prod_{k=0}^{n-1}\sigma_k \hat s_k~,}
is constructed out of the {\it chiral} operators $\sigma_k\hat s_k$. Indeed, they have dimension \dimsigma, \dimshatk\ $h(\sigma_k\hat s_k)=\half {k\over n}$ and R-charge \uone\ $R_k={k\over n}$, so \hhh\ is satisfied. The resulting dimension and R-charge of $\CS_n$ are
\eqn\hhssnn{h(\CS_n)=\half R(\CS_n)={n-1\over 4}~,}
strictly larger than \hhttnn, as expected.

We can now consider the correlation function
\eqn\oonn{\CO_n=\langle \CS_n(u_1)\CS_{n}^*(v_1)\CS_n(u_2)\CS_{n}^*(v_2)\cdots \CS_n(u_N)\CS_{n}^*(v_N)\rangle~,}
which is obtained from \fff\ by replacing the usual twist fields $\CT_n$ with their chiral analogs. The supersymmetric Renyi entropy is obtained by writing the analog of \bbb,
\eqn\susyrenyi{S_A^{(n)}({\rm susy})={1\over 1-n}\ln\CO_n~.}
For one interval, one has (in similar notation to \ggg)
\eqn\oneinterval{\CO_n=\langle\CS_n(u)\CS_{n}^*(v)\rangle\sim l^{-4h(\CS_n)}~,}
which leads to the SRE
\eqn\susyrenyifree{S_A^{(n)}({\rm susy})=\ln{l\over a}~.}
We see that
\item{(a)} The SRE is independent of $n$ in this case.
\item{(b)} It is equal to the standard entanglement entropy \oneint\ (with $c=3$).

\noindent
As we will see below, property (a) is  special to the one interval case, but (b) is a special case of the more general statement that
\eqn\linsusyren{\lim_{n\to 1} S_A^{(n)}({{\rm susy}})=S_A~,}
\ie\ the supersymmetric Renyi entropy gives the standard entanglement entropy in the limit $n\to 1$.

For the general case of $N$ intervals, the two Renyi entropies defined via \bbb, \fff, and \oonn, \susyrenyi, are difficult to compute. However, as noted above, the complicated part is the $2N$ point function of bosonic twist operators $\sigma_k$, which is the same in the two cases. Therefore, when computing the difference
\eqn\diffent{S_A^{(n)}({\rm susy})-S_A^{(n)}={1\over 1-n}\ln {\langle \CS_n(u_1)\CS_{n}^*(v_1)\CS_n(u_2)\CS_{n}^*(v_2)\cdots \CS_n(u_N)\CS_{n}^*(v_N)\rangle \over
\langle \CT_n(u_1)\CT_{n}^*(v_1)\CT_n(u_2)\CT_{n}^*(v_2)\cdots \CT_n(u_N)\CT_{n}^*(v_N)\rangle}~,}
it cancels. The ratio of correlation functions in \diffent\ only involves the fermionic twist fields  \fermtwist, \susyfermtwist, and thus can be computed exactly using free field techniques. One finds (after adding the right-moving contributions to $\CS_n$, $\CT_n$; see footnote 5)
\eqn\ratiosshat{{\langle \CS_n(u_1)\CS_{n}^*(v_1)\cdots \CS_n(u_N)\CS_{n}^*(v_N)\rangle \over
\langle \CT_n(u_1)\CT_{n}^*(v_1)\cdots \CT_n(u_N)\CT_{n}^*(v_N)\rangle}=\left|\prod_{i<j}(u_i-u_j)(v_i-v_j)\prod_{i,j}(v_j-u_i)^{-1}\right|^{4[h(\CS_n)-h(\CT_n)]}~.
}
Looking back at \hhttnn, \hhssnn, we see that
\eqn\diffh{h(\CS_n)-h(\CT_n)={(n-1)^2\over 8n}~.}
Plugging this into \diffent\ we finally have
\eqn\diffrenyient{S_A^{(n)}({\rm susy})-S_A^{(n)}={1-n\over 2n}\left[\sum_{1=i<j}^N\left(\ln|u_i-u_j|+\ln|v_i-v_j|\right)-\sum_{i,j=1}^N\ln |v_j-u_i|\right]~.}
We see that
\item{(a)} While each of the two Renyi entropies on the l.h.s. of \diffrenyient\ is complicated, the difference between them is simple.
\item{(b)} As $n\to 1$, $S_A^{(n)}$ approaches the entanglement entropy \ccc\ for the region \ddd, and according to \diffrenyient\ so does $S_A^{(n)}(\rm susy)$.

In the next section we will see that the discussion of this section can be extended to a general $(2,2)$ SCFT. For the purpose of this generalization it is useful to present the results of this section in a more algebraic way.

As mentioned above, the twist operator $\CT_n$ \formttnn\ has vanishing charge under the $U(1)_R$ current of $\CM^n/\IZ_n$, \uone. In fact, it satisfies the stronger property that its OPE with $J$ is regular, \ie\ it belongs to the coset $\left(\CM^n/\IZ_n\right)/U(1)_R$. This means that if we define a canonically normalized scalar field $H$ via $J=i\sqrt{n}\partial H$, and consider the operators
\eqn\tnalpha{\CT_n(\alpha)=\CT_ne^{i\alpha H}~,}
their dimension and R-charge are given by
\eqn\hrtnalpha{h(\CT_n(\alpha))=h(\CT_n)+\half\alpha^2~;\qquad R(\CT_n(\alpha))=\alpha\sqrt n~.
}
It is natural to ask whether we can adjust $\alpha$ such that the operator $\CT_n(\alpha)$ satisfies the chirality condition \hhh. A short calculation leads to\foot{There is another solution that does not have the right $n\to 1$ limit.}
\eqn\formalal{\alpha=\alpha_n={n-1\over2\sqrt n}~.}
The resulting R-charge and dimension are precisely those of $\CS_n$, \hhssnn. In fact, it is easy to check directly that
\eqn\reltnsn{\CS_n=\CT_n e^{i\alpha_nH}~.}
Thus, the chiral operator obtained from the construction of \tnalpha\ -- \formalal\ is precisely the one that gives rise to the SRE \oonn, \susyrenyi. As we will see in the next section, the decomposition \reltnsn\ leads directly to \diffrenyient.

\newsec{The general case}

In this section we extend the construction of  the previous section to a general $\CN=(2,2)$ SCFT $\CM$ with central charge $c$. The Renyi entropy of this theory is again obtained by studying the cyclic orbifold $\CM^n/\IZ_n$. The lowest dimension operator in the $\IZ_n$ twisted sector, $\CT_n$ \fff, has in this case the scaling dimension \eee\ and it commutes with the $U(1)_R$ current
\eqn\genuoner{J=\sum_{j=1}^n J_j~,}
as in the free field case analyzed in section 2. We can thus generalize the construction described at the end of section 2 to this case. We define a canonically normalized scalar field $H$ via
\eqn\uoner{J=i\sqrt{cn\over 3}\partial H~,}
and consider the operator $\CT_n(\alpha)$ \tnalpha, whose dimension is given by \hrtnalpha, while the R-charge is now
\eqn\rtnalgen{R(\CT_n(\alpha))=\alpha\sqrt{cn\over 3}~.}
As there, we can tune $\alpha$ such that the operator $\CT_n(\alpha)$ is chiral.  This leads to
\eqn\formalalgen{\alpha=\alpha_n={n-1\over2}\sqrt{c\over3n}~.}
The corresponding operator,
\eqn\defsngen{\CS_n=\CT_n e^{i\alpha_nH}~,}
is the chiral operator whose correlation functions determine the SRE via \oonn, \susyrenyi. Its dimension and R-charge are given by
\eqn\hrchiral{h({\cal S}_n)={1\over 2}R(\CS_n)={c\over 12}(n-1)~.}
Note that this is in agreement with the chiral spectrum of symmetric products as analyzed \eg\ in section 5.1 of  \ArgurioTB.

For one interval, one finds the analog of \susyrenyifree,
\eqn\susyrenyifree{S_A^{(n)}({\rm susy})={c\over 3}\ln{l\over a}~,}
which is again independent of $n$ and equal to the usual entanglement entropy $S_A$ given by \oneint. For an arbitrary number of intervals, one needs to generalize the calculation \ratiosshat. The key point is that the numerator of that expression splits into two decoupled factors,
\eqn\splitcorr{\langle \CS_n(u_1)\cdots \CS_{n}^*(v_N)\rangle=\langle \CT_n(u_1)\cdots \CT_{n}^*(v_N)\rangle
\langle\prod_{j=1}^Ne^{i\alpha_n H(u_j)}e^{-i\alpha_n H(v_j)}\rangle~.
}
The first is the contribution to the correlator of $\left(\CM^n/\IZ_n\right)/U(1)_R$, the second is the $U(1)_R$ CFT contribution.  The ratio \diffent\ receives contributions only from the latter, and is given again by \ratiosshat. The difference of dimensions takes in this case the form
\eqn\diffhgen{h(\CS_n)-h(\CT_n)={c\over 24n}(n-1)^2~,}
which reduces to \diffh\ for $c=3$. The corresponding difference of entropies is
\eqn\diffrenyientgen{S_A^{(n)}({\rm susy})-S_A^{(n)}={c\over 6}{1-n\over n}\left[\sum_{1=i<j}^N\left(\ln|u_i-u_j|+\ln|v_i-v_j|\right)-\sum_{i,j=1}^N\ln |v_j-u_i|\right]~.}
As in section 2, we see that the difference between the two Renyi entropies goes to zero as $n\to 1$, and for general $n$ they differ by a simple universal\foot{In the sense that it is independent of any details of the SCFT $\CM$.} function of the parameters $(c,n, u_i, v_i)$.

\newsec{String theory on $AdS_3$}

A large class of two dimensional CFT's is obtained by studying string theory on $AdS_3$.  In this case, we can make use of  the fact that the worldsheet CFT on the $AdS_3$ background (supported by Neveu-Schwarz $B_{\mu\nu}$ field) can be described as a WZW model on the $SL(2,\IR)$ group manifold. To get theories with $(2,2)$ superconformal symmetry, it is natural to consider string backgrounds of the form
\eqn\twotwost{AdS_3\times S^1\times \CN~,}
where $\CN$ is a worldsheet SCFT with $(2,2)$ superconformal symmetry.\foot{Not to be confused with the $(2,2)$ superconformal symmetry of the spacetime, or boundary, SCFT, which is our focus here.}
The level $k$ of the $SL(2,\IR)$ WZW model is related to the worldsheet central charge of $\CN$; see \eg\ \GiveonNS\ for the details.  In this class of models one can impose a chiral GSO projection that leads to $(2,2)$ superconformal symmetry in spacetime
\refs{\GiveonJG,\BerensteinGJ}.

It is natural to ask how to calculate the Renyi entropy of the spacetime SCFT in these models. In this section we will argue that string theory on \twotwost\ computes the {\it supersymmetric} Renyi entropy discussed in the previous sections. A useful observation for that purpose, \ArgurioTB, is that the spectrum of chiral operators in the spacetime SCFT is the same as that on the symmetric product \symprod. Here $\CM$ is a $(2,2)$ SCFT with central charge $c=6k$,
whose properties are implicitly determined by the worldsheet SCFT $\CN$ in \twotwost, and $p$ is an integer related to the string coupling, $p\sim 1/g_s^2$. For example, for string theory on $AdS_3\times S^3\times T^4$, which can be viewed as a special case of \twotwost\ (after imposing a chiral GSO projection in the latter), one has $\CM=T^{4k}/S_k$; more precisely, $\CM$ is in the moduli space of this symmetric orbifold.

Based on this and other observations, the authors of \ArgurioTB\ proposed that the spacetime SCFT corresponding to \twotwost\ {\it is}\foot{Actually, \ArgurioTB\ proposed that it is in the moduli space of the symmetric product, but in general this symmetric product does not have any marginal operators, so if the proposal is correct, it is the symmetric product.} the symmetric product \symprod. If this is correct, the only non-trivial Renyi entropy one needs to compute is that for $\CM$. Thus, one needs to identify in string theory on $AdS_3$ the appropriate twist operators.

There is a very natural candidate for the chiral operators $\CS_n$. As discussed in
\ArgurioTB, chiral operators in the $\IZ_n$ twisted sector of the spacetime SCFT \symprod\ are described by vertex operators with winding $n$, in the sense of \MaldacenaHW.
In particular, starting with a chiral operator with spacetime dimension $h_1$ in the short string sector, one can construct  a sequence of chiral operators in sectors with winding $n$, whose dimensions are given by
\eqn\hnhone{h_n={1\over 2}R_n=h_1+{c\over 12}(n-1)~,\qquad n=1,2,\cdots~.}
For the special case $h_1=0$ this leads to an operator with the same quantum numbers as $\CS_n$ \hrchiral, and it is natural to identify the two.
Computing the correlation functions \oonn\ of these operators thus computes the SRE of $\CM$.

Physical vertex operators corresponding to chiral operators in the untwisted sector of the spacetime orbifold
are obtained by dressing the chiral operators in the internal worldsheet
theory $\CN$ by an appropriate factor from the $AdS_3\times S^1$ piece
\refs{\GiveonNS,\KutasovZH,\ArgurioTB}.
The vertex operators corresponding
to chiral operators in the $\IZ_n$ twisted sectors of the spacetime theory
are obtained by acting on the untwisted ones with a simple
`chiral twist' operator in the $SL(2,\IR)\times U(1)$ worldsheet theory;
see \ArgurioTB\ for the detailed construction.

In particular, the spacetime identity operator, $\CS_1\equiv I$ \refs{\KutasovXU,\PorratiEHA}, is obtained from the identity in $\CN$.
Consequently, the vertex operators for $\CS_n$ discussed above live purely in the $AdS_3\times S^1$ factor, and thus do not contain any information about the particular SCFT $\CN$ in \twotwost\ that enters the construction, except for the level $k$, which determines the spacetime central charge of $\CM$. Thus, any SCFT with a description in terms of string theory on $AdS_3$ with NS $B$ field has the property that its Renyi entropy is universal, at least to leading order in $g_s^2\simeq 1/p$.

\newsec{Discussion}

In this note we showed that in any two dimensional conformal field theory with $(2,2)$ SUSY one can obtain the Renyi entropy \bbb\ for any spatial region $A$ \ddd\ by calculating a correlation function of chiral and anti-chiral operators.  This was done by defining the chiral operators $\CS_n$ \defsngen, and showing that their correlation function \oonn, which leads to the supersymmetric Renyi entropy \susyrenyi, satisfies the relation \diffrenyientgen\ and thus determines the Renyi entropy $S_A^{(n)}$.

We further showed that the operators $\CS_n$ are present in string theory on $AdS_3$ with $(2,2)$ superconformal symmetry in spacetime, and thus in any $(2,2)$ SCFT with an $AdS_3$ string dual one can compute the correlators \oonn, and hence the Renyi entropy \bbb. The resulting entropy is universal -- the only data it depends on is the parameters $(k,n, u_i, v_i)$. In particular, it is independent of the detailed structure of the string background  \twotwost.

There is a number of natural extensions of our results. One is to use the chirality of the operators $\CS_n$ to compute the correlation functions \oonn, and thus shed new light on the Renyi entropy. Another is to calculate these correlation functions in string theory on $AdS_3$ using worldsheet techniques. It would also be interesting to generalize the discussion to the case of RG flows that preserve $(2,2)$ supersymmetry.

Another interesting direction involves the relation of our results to \NishiokaHAA. In that paper the authors defined a supersymmetric version of Renyi entropy in three dimensional  theories with $\CN=2$ SUSY via partition functions on branched covers of the three-sphere. This construction has been further developed in \HuangGCA, and generalized to four \refs{\HuangPDA,\CrossleyOEA} and five \HamaIEA\ dimensions; its gravity dual was discussed in \refs{\HuangGCA,\NishiokaMWA,\CrossleyOEA,\AldayFSA}. It would be interesting to generalize this work to two dimensions, and see whether it gives the same quantity as our supersymmetric Renyi entropy. If that's the case, it would be interesting to understand our result \diffrenyientgen\ from this perspective, and study possible generalizations of this result to higher dimensions.

\bigskip\bigskip

\noindent{\bf Acknowledgements:}
The work of AG is supported in part by the I-CORE Program of the Planning and Budgeting Committee and the Israel Science Foundation (Center No. 1937/12), and by a center of excellence supported by the Israel Science Foundation (grant number 1989/14). DK is supported in part by DOE grant DE FG02-13ER41958. DK thanks Tel Aviv University and the Hebrew University for hospitality during part of this work.

\listrefs

\end